\documentclass[twocolumn]{aastex63}
\usepackage{amsmath,amstext,amssymb}

\usepackage{amsmath,amstext,amssymb,amsfonts}
\usepackage{epsfig}
\usepackage{url}
\usepackage{hyperref}

\usepackage[normalem]{ulem} 

\usepackage{latexsym}
\usepackage{epsfig}
\usepackage{wasysym}
\usepackage{graphicx}
\usepackage{verbatim}
\usepackage{tikz} 

\newcommand{\Milwaukee}{University of Wisconsin-Milwaukee, Milwaukee, WI 53201, USA}

\shorttitle{GW polarizations from strong lensing}
\shortauthors{Maga\~na~Hernandez}

\graphicspath{{./}{figures/}}

\begin{document}

\title{Measuring the polarization content of gravitational waves with strongly lensed binary black hole mergers}

\correspondingauthor{I. Maga\~na~Hernandez}
\email{maganah2@uwm.edu}
\author[0000-0003-2362-0459]{I. Maga\~na~Hernandez}
\affiliation{\Milwaukee}

\begin{abstract}
Alternative theories of gravity predict up to six distinct polarization modes for gravitational waves. Strong gravitational lensing of gravitational waves allows us to probe the polarization content of these signals by effectively increasing the number of observations from the same astrophysical source. The lensing time delays due to the multiple observed lensed images combined with the rotation of the Earth allows for effective non-collocated interferometers to be defined with respect to the source location and hence probe the alternative polarization amplitudes with more observations. To measure these amplitudes, we jointly fit the image observations to a single gravitational wave signal model that takes into account the image magnifications, time delays, and polarization mode amplitudes. We show that for certain systems, we can make a measurement of the relative mode amplitudes for lensed events with two detectable images.
\end{abstract}

\section{Introduction}
\label{sec:intro}
The latest set of gravitational--wave (GW) observations released by the LIGO Scientific \citep{LIGOScientific:2014pky}, Virgo \citep{VIRGO:2014yos} and KAGRA \citep{Aso:2013eba} Collaboration (LVK) as part of The third Gravitational-wave Transient Catalog (GWTC-3) catalog \citep{LIGOScientific:2018mvr,LIGOScientific:2020ibl,LIGOScientific:2021usb,LIGOScientific:2021djp} contains 69 confident binary black hole (BBH) detections as well as both confident detections for binary neutron star and neutron star black hole mergers. As a consequence, the increasing size of gravitational wave catalogs has allowed for in-depth studies of the binary black hole population properties \citep{LIGOScientific:2018jsj,LIGOScientific:2020kqk,LIGOScientific:2021psn}, cosmic expansion history \citep{LIGOScientific:2019zcs,LIGOScientific:2021aug} as well as tests of general relativity in the strong field regime \citep{LIGOScientific:2019fpa,LIGOScientific:2020tif,LIGOScientific:2021sio} including a search for gravitational wave lensing signatures \citep{LIGOScientific:2021izm}.

When gravitational waves propagate and interact with intervening matter such as galaxies or dense galaxy clusters, there is a change for strong gravitational lensing and for multiply lensed GW images to be produced with time delays ranging from minutes to months \citep{Takahashi:2003ix,Haris:2018vmn,Dai:2016igl}. Over the upcoming years, ground-based GW detectors such as Advanced LIGO, Advanced Virgo and KAGRA are expected to find 0.1 to 1 pairs of strongly lensed GW signals per year originating from binary black hole mergers at their corresponding design sensitivities \citep{Ng:2017yiu,LIGOScientific:2021izm,Xu:2021bfn,Caliskan:2022wbh,Mukherjee:2020tvr}. In fact, the first search for signatures of lensing (including strongly lensed pairs) was performed in \cite{Hannuksela:2019kle} using the 10 BBH events of the GWTC-1 catalog \citep{LIGOScientific:2018mvr}. No conclusive evidence for a strongly lensed pair was found, however, the pair with the highest evidence favoring the lensing hypothesis was GW170104/GW170814 as pointed in \cite{Hannuksela:2019kle,McIsaac:2019use}. Subsequent studies followed up the pair with a fully Bayesian joint parameter estimation study over the lensed images and arrived at similar conclusions disfavoring the lensing hypothesis \citep{Liu:2020par,Dai:2020tpj}. The most comprehensive study to date using the first half of LIGO-Virgo's third observation run observations has also yielded no substantial evidence for lensing \citep{LIGOScientific:2021djp,LIGOScientific:2021izm}

Alternative metric theories of gravity predict up to six distinct polarization modes for GW emission, besides the two tensorial modes allowed by general relativity \citep{Isi:2017equ,Chatziioannou:2012rf}. In order to probe the presence (or lack off) for these alternative polarizations, a network of six linearly independent detectors is needed. Future ground based detector networks will allow for some statements about the relative amplitudes for each mode, however, discerning the full polarization content would be difficult for most systems \cite{Chatziioannou:2021mij}. The most recent observational results using the full GWTC-3 catalog have placed stringent constrains on alternative polarizations being present \citep{LIGOScientific:2019fpa,LIGOScientific:2020tif,LIGOScientific:2021sio}. The strongest of such constraints disfavour the presence of vector or scalar modes being present individually when compared to the expected GR tensor modes. However, the presence of tensor modes as well as either vector or scalar modes (or both) as a fully mixed model has yet to be constrained strongly. 

In this work, we explore constraints on alternative GW polarizations with simulated pairs of strongly lensed GW signals. We parameterize the GW model as a fully mixed tensor, vector and scalar mode model with up to 5 degrees of freedom allowing us to make statements about the relative amplitudes for each mode. The difference in arrival times for each strongly lensed image probes the same GW signal arriving at different times. The rotation of the Earth imposes the time dependence of the antenna beam pattern functions, allowing us to see a different projection for the GW signal at each detector (essentially doubling the number of detectors for a pair of lensed events) \citep{Goyal:2020bkm}. We measure the relative amplitudes for each mode by jointly fitting the detected lensed image pairs using the framework described in \citep{Liu:2020par,Lo:2021nae} and show that for some systems the polarization mode amplitude degeneracies can be broken with a single pair of lensed events.

This paper is organized as follows. In Section 2 we describe the alternative (non-tensorial) polarization modes for gravitational-wave signals. In Section 3, we summarize the effect of strong lensing in detected GW signals, focusing on pairs of lensed events. In Section 4, we present the main results of this paper and in Section 5, we provide a summary of this work. We use the Planck 2015 cosmological model \citep{Planck:2015fie} throughout this paper, that is, $H_0 = 67.8 \ \rm km/s/Mpc$, $\Omega_{m} = 0.308$, $\Omega_{\Lambda} = 0.692$ and set $\Omega_{k,0} = 0$.

\section{Nontensor polarizations}
\label{sec:nongr}
\newcommand{\mplus}{\ensuremath{+}}
\newcommand{\mcross}{\ensuremath{\times}}
\newcommand{\mx}{\ensuremath{x}}
\newcommand{\my}{\ensuremath{y}}
\newcommand{\mb}{\ensuremath{b}}
\newcommand{\ml}{\ensuremath{l}}
\newcommand{\ms}{\ensuremath{s}}

\newcommand{\hplus}{\ensuremath{h_+}}
\newcommand{\hcross}{\ensuremath{h_\times}}
\newcommand{\hx}{\ensuremath{h_x}}
\newcommand{\hy}{\ensuremath{h_y}}
\newcommand{\hb}{\ensuremath{h_b}}
\newcommand{\hl}{\ensuremath{h_l}}
\newcommand{\hs}{\ensuremath{h_s}}

Alternative metric theories of gravity (beyond general relativity) may allow up to six distinct polarization modes on the GW waveform, including the two tensor $\mplus$ and $\mcross$ modes expected in general relativity (GR). These additional polarization modes are the two vector modes $\mx$ and $\my$, as well as two scalar modes $\mb$ and $\ml$ (breathing and longitudinal respectively). The GW perturbation can thus be written as,
\begin{equation} \label{eq:h_A}
h_{ij} = \sum_A h_A\, e^A_{ij} \, ,
\end{equation}
where $e^A_{ij}$ is the polarization tensor for mode $A$ and $h_A$ are the corresponding polarization mode amplitudes. The GW perturbation is thus a linearly independent weighted sum over modes, the most generic case corresponding to $A \in \{\mplus,\mcross, \mx, \my, \mb, \ml \}$. 

In general, GW interferometers measure the projection of the perturbation given by Eq.~\eqref{eq:h_A} onto the detector arms. Thus the measured GW strain at detector $I$ can be written as,
\begin{equation} 
h_I(t) = \sum_A F_I^A(\alpha, \delta, \psi, t)\, h_A(t)\, ,
\end{equation}
with antenna beam pattern functions $F_I^A \equiv D_I^{ij} e^A_{ij}$ defined with respect to the detector tensor $D_I^{ij}$ which encodes the geometry of the GW detector. The antenna pattern functions are in general functions of time and depend on the sky location of the GW source defined by its right ascension $\alpha$ and declination $\delta$ as well as the polarization angle $\psi$. It is worth noting that the breathing and longitudinal mode antenna pattern functions are identical (up to a constant) so that $F_\mb = - F_\ml$. This degeneracy, makes each scalar mode contribution difficult to disentangle unless a specific modified theory of gravity is chosen a-priori, leading to model dependent constraints. Following convention we pick the breathing mode as the scalar mode of interest, thus the sum over linearly independent modes in Eq.~\eqref{eq:h_A} reduces to a sum over five polarization modes, $A \in \{\mplus,\mcross, \mx, \my, \ms\}$ where we denoted the breathing mode ($\mb$) by ($\ms$) for convenience. For a detailed discussion on GW polarizations and the various polarization angle conventions we refer the reader to \citep{Isi:2017equ,Isi:2022mbx}.

\begin{figure}
\centering
\includegraphics[width=0.45\textwidth]{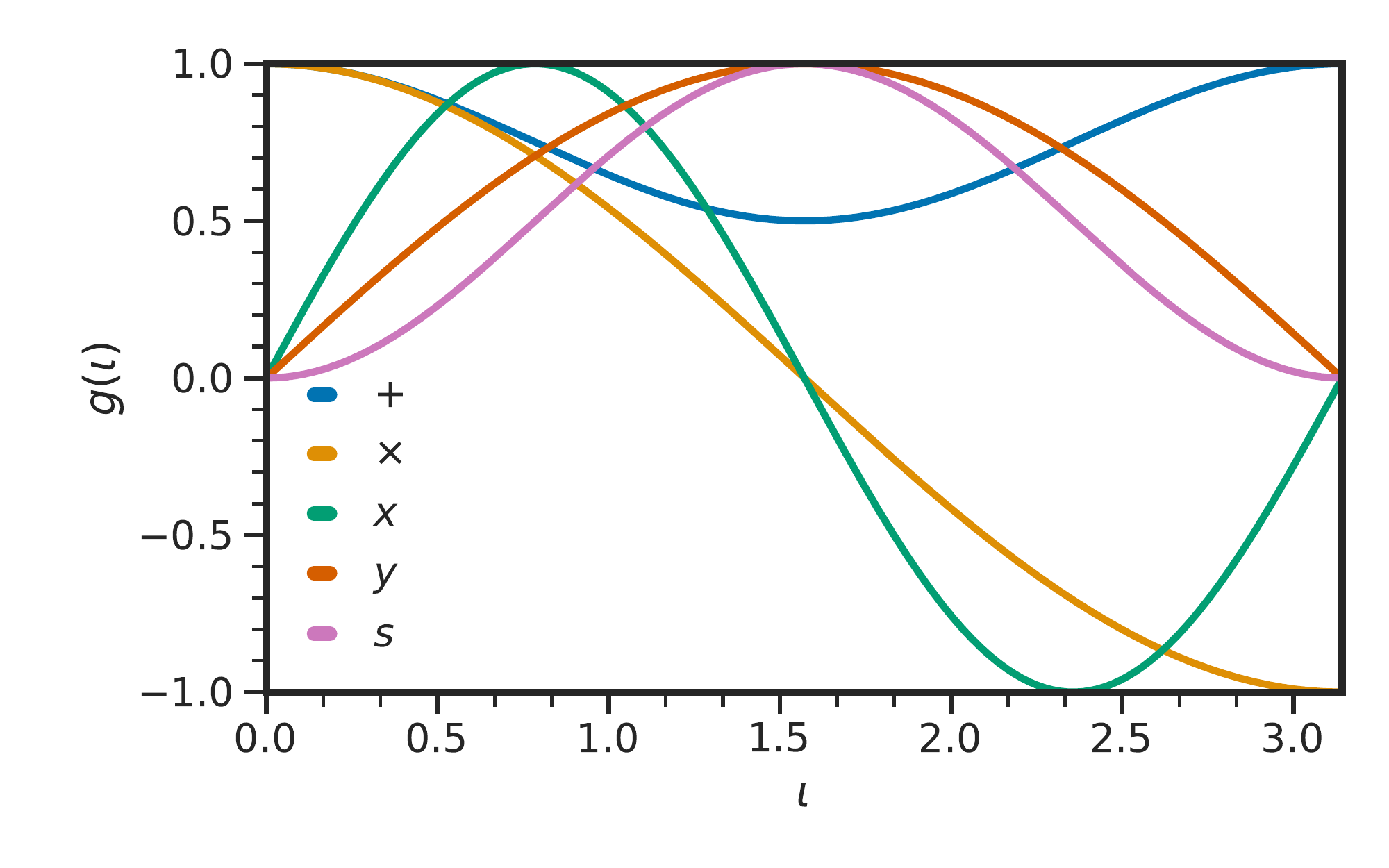}
\caption{\label {fig:g_iota} We show the values for $g(\iota)$ as a function of the inclination angle $\iota$ for each of the six polarization modes. We note that for face-on systems ($\iota=0$), the vector and scalar modes will not be present in the GW strain data even if emitted. For edge-on systems ($\iota=\pi/2$), $\times$-mode and the vector-$x$ will not be detectable. The optimal inclination angle for which we maximize over the presence of all modes in the GW data corresponds to  $\iota_\text{opt} \approx 0.87$. }
\end{figure}

For gravitational waves produced by a compact binary merger such as a pair of merging binary black holes, there is an additional inclination angle dependence for each polarization mode \citep{Chatziioannou:2012rf,Takeda:2020tjj}. We define this dependence via the function $g_A(\iota)$, so that $g_\mplus(\iota) = (1+\cos^2\iota)/2$, $g_\mcross(\iota) = \cos\iota$, $g_\mx(\iota) = \sin{2\iota}$,  $g_\my(\iota) = \sin\iota$ and  $g_{\mb,\ml}(\iota) = \sin^2\iota$ where $\iota$ is the inclination angle of the binary. We can thus write the gravitational wave strain at detector $I$ as, 
\begin{equation} \label{eq:h_measured}
h_I(t) = \sum_A F_I^A(\alpha, \delta, \psi, t)\, g_A(\iota)\, h_A(t)\,.
\end{equation}
From the above expression, we can see that the inclination angle dependence on the polarization modes is important since for a face-on system ($\iota=0$) only the tensor modes will be present in the data while for an edge-on system ($\iota=\pi/2)$, only the cross polarization mode vanishes but all other modes are present. The inclination angle dependence is critical for 2nd generation ground based detectors since we expect most mergers to be near the face-on limit. In Fig. \ref{fig:g_iota}, we plot the dependence on inclination for the mode amplitudes. Clearly, if all other parameters are fixed, then the optimal inclination would be $\iota_\text{opt} \approx 0.87$.

\section{Strong Gravitational Wave Lensing}
As gravitational waves propagate, there is a chance for strong gravitational lensing to occur due to intervening galaxies or larger cosmic structures such as galaxy clusters. The strong lensing of gravitational waves can give rise to multiple images of the same GW transient each with its own absolute magnification factor $\mu_k$. When the GW images are detected, each will arrive at a different time $t_c^{(k)}$ and each might have a frequency independent phase shift (Morse phase) $\Delta\phi_k = -\pi n_k /2$ with index $n_k = 0, 1, 2$ defining Type-I, Type-II and Type-III images respectively. The gravitational wave waveform for each lensed image is then given by,
\begin{equation} \label{eq:h_lens}
    h_L(f,\theta,\mu_k, t_c^{(k)},\Delta\phi_k) = \sqrt{\mu_k}\exp \left(if\Delta\phi_k\right)h_U(f, \theta, t_c^{(k)})
\end{equation}
where $h_U$ is the waveform without any strong lensing effects (unlensed) and $\theta = \{m_1, m_2, a_1, a_2, \iota, \alpha, \delta, \psi\}$ where $m_1$ and $m_2$ are the primary and secondary masses of the binary in the source frame, $a_1$ and $a_2$ are the (aligned) component spin magnitudes. The set of parameters $\theta$ is common across all lensed images, including the sky location of the GW source due to expected order of arcsecond deflection angles for each image being much smaller than the typical localization regions for 2G detectors \cite{Takahashi:2003ix}.

Now, for a source at luminosity distance $D_L$, the lensed images are magnified (de-magnified) by their corresponding magnification factors  $\sqrt{\mu_k}$ as in Eq.~\eqref{eq:h_lens} so that the the observed distances correspond to,
\begin{equation}
D^{(k)}_{\text{obs}}=D_L/\sqrt{\mu_k}\,,
\end{equation}
clearly showing the degeneracy between the luminosity distance to the source and the absolute magnification factors for each lensed image.
For a pair of lensed images, it is convenient to define the relative magnification factor $\mu$ as,
\begin{equation}
\mu = \left( \frac{D^{(1)}_{\text{obs}}}{D^{(2)}_{\text{obs}}} \right)^2 = \frac{\mu_2}{\mu_1} \,,
\end{equation}
where we label the signal that is detected first by $(1)$ and consequently the later arriving signal by $(2)$. Finally we can also define the lensing time delay for the pair of lensed images as $\Delta t = t_c^{(2)} - t_c^{(1)}$ which is always greater than zero.

\begin{figure*}[htb]
\includegraphics[width=0.45\textwidth]{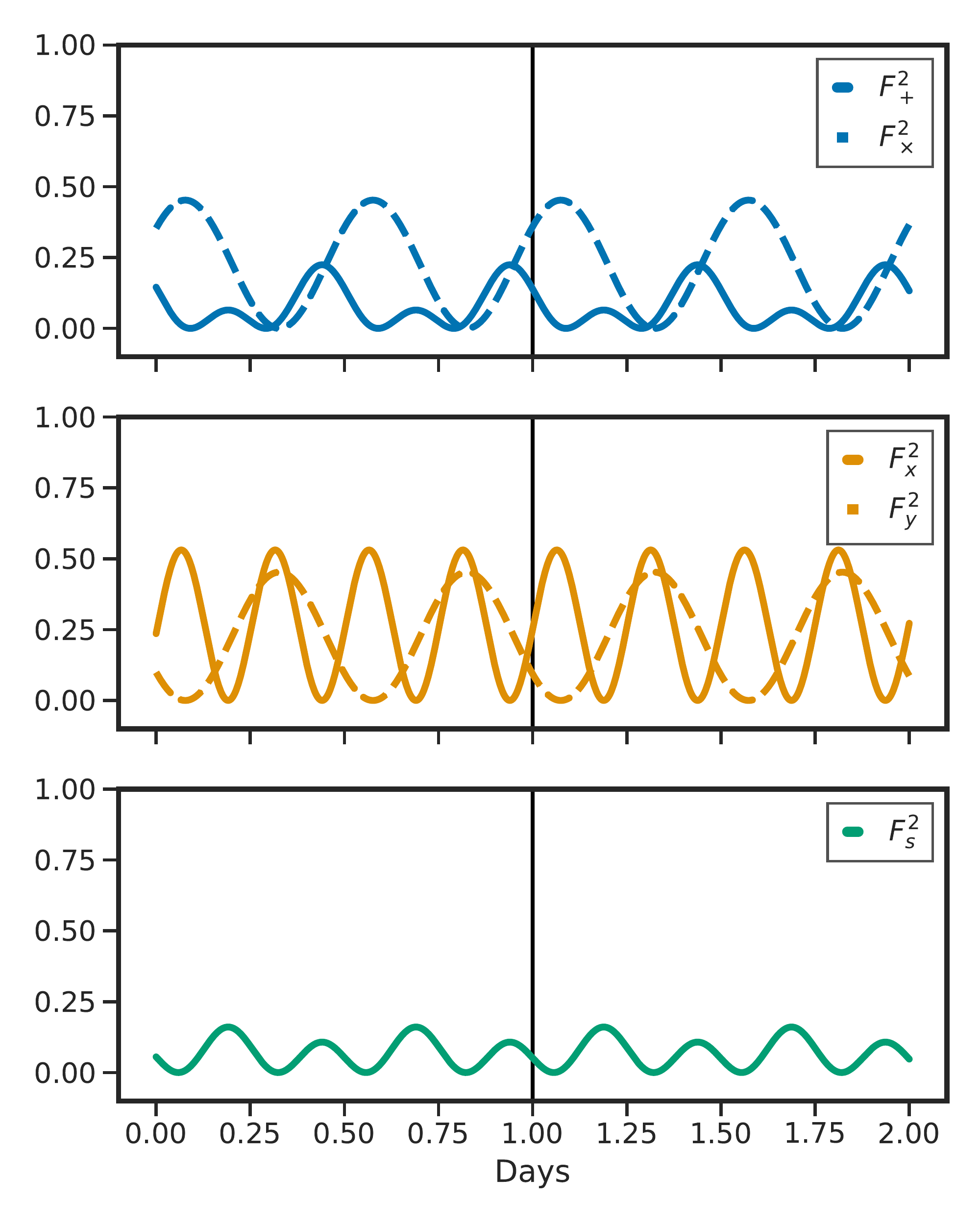}
\includegraphics[width=0.45\textwidth]{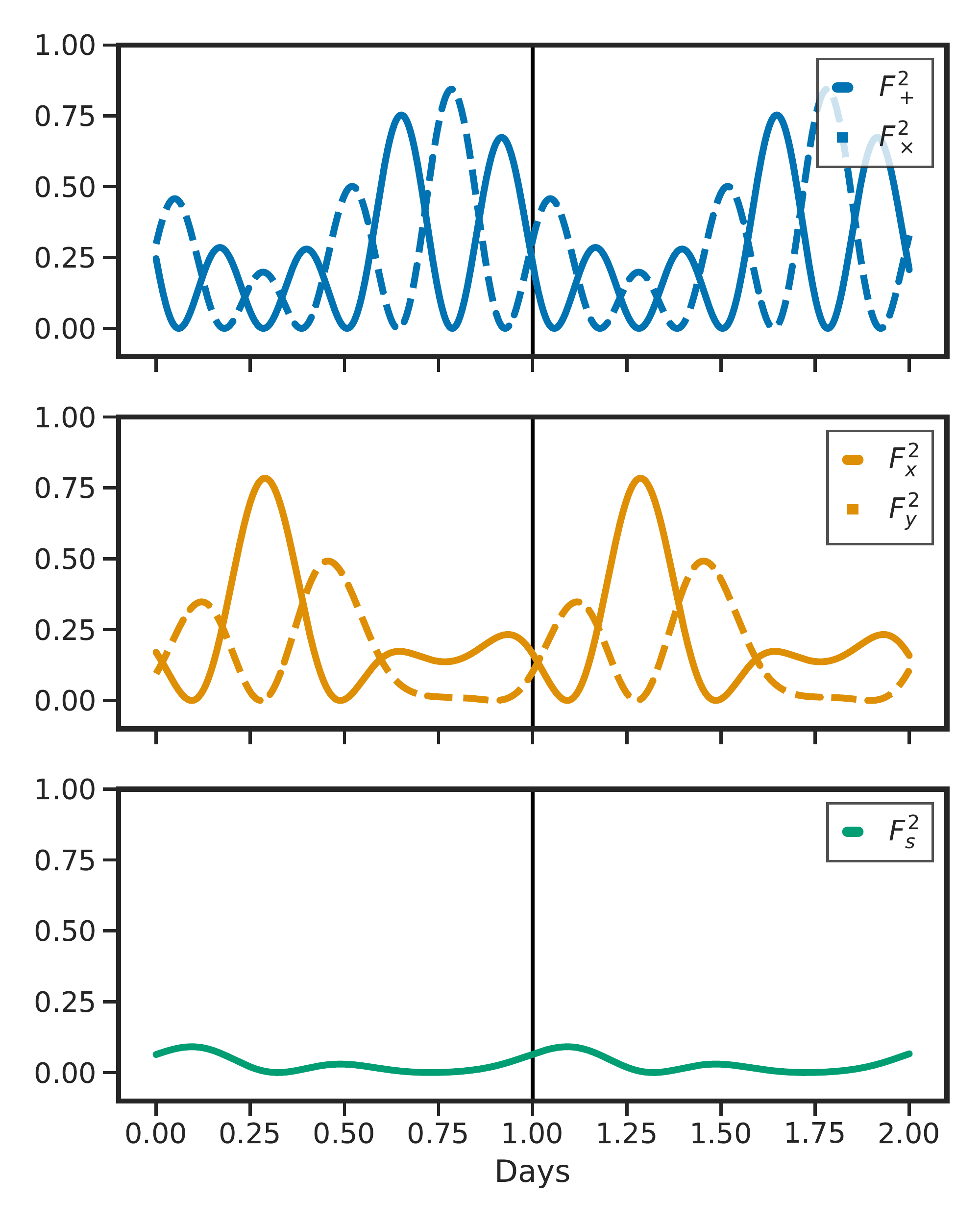}
\caption{\label {fig:antenna} We show the values for $F_A^2(t)$ for the six polarization modes over the span of two days where we have fixed the polarization angle to $\psi=0$ for convenience. In the left panel we show an example where the sky location of the source is fixed at $(\alpha,\delta) = (0,0)$. Similarly, we show another example but with $(\alpha,\delta) = (1.375,-1.211)$ to illustrate the complex behavior of the antenna beam pattern functions in terms of sky location and time.}
\end{figure*}

The most important effect due to strong lensing and the production of multiple images is the different times of arrival for each lensed image. Due to Earth's rotation the location of the network of detectors will change as a function of time relative to the sky location of the lensed signals. This means that the antenna pattern functions for each polarization mode will probe the polarization content of the arriving GW signal differently depending on the arrival time of each lensed image, allowing us to constrain the relative amplitudes for each polarization mode \cite{Goyal:2020bkm}. In principle, this leads to effectively doubling the number of detectors in the network for a pair of strongly lensed images. In order to illustrate this point we show the antenna pattern functions $F_A^2(t)$ in Fig. \ref{fig:antenna} at a fixed polarization angle $\psi=0$ for two different sky locations over a period of two days. As mentioned in Section \ref{sec:intro}, the expected time delay between a pair of lensed events by an intervening galaxy could range from hours to months. With respect to probing the polarization amplitudes for each mode, the expected time delay is not important but the relative time delay corresponding to the rotation of the Earth over a day.  

\section{Joint Parameter Estimation}
\label{sec:pe}
Since strongly lensed systems leave the frequency evolution of the gravitational-wave binary unchanged and thus only induce an overall amplitude and phase difference amongst the detected images. We are thus able to jointly fit the lensed events by taking into account the predicted strong lensing effects on the GW-waveform. We provide a summary for the joint parameter estimation below. For the full derivations and detailed discussion of the framework, see \citep{Liu:2020par,Lo:2021nae}.

Under the assumption that we have a confidently detected pair of strongly lensed GW events. We can perform joint parameter estimation by considering the strong lensing waveform model for each detected image in Eq.~\eqref{eq:h_lens} and use the model with alternative polarizations as defined in Eq.~\eqref{eq:h_measured} as the definition for $h_U(t,\theta)$. Additionally we parameterize each polarization mode amplitude by a set of relative amplitude parameters $\{\epsilon_A\}$ which must satisfy the following constraint $\sum_A \epsilon_A = 1$. For GR, we must have $\epsilon_\mplus = \epsilon_\mcross = 0.5$ while the vector and scalar mode contributions are all zero. 

Under the lensing hypothesis, for a pair of lensed events with measured strains $d_1$ and $d_2$, we jointly infer the binary parameters $\theta$, the lensing observables $\{\mu, \Delta t\}$ (in this work we set $\Delta\phi_k=0$ for all images) as well as the relative amplitudes for each polarization mode $\{\epsilon_A\}$ in terms of the observed distance and time of arrival of the first image, 
\begin{align} \label{eq:likelihood}
\begin{split}
    &\mathcal{L}(d_1, d_2 |\theta, D^{(1)}_{\text{obs}}, t_c^{(1)}, \mu, \delta t, \{\epsilon_A\} ) \\
    &=\mathcal{L}(d_1|\theta, D^{(1)}_{\text{obs}}, t_c^{(1)}, \{\epsilon_A\} ) \mathcal{L}(d_2 |\theta, \mu, \delta t, \{\epsilon_A\} ),
\end{split}
\end{align}
where $\mathcal{L}(d_1, d_2 | \ldots)$ is referred to as the strong lensing joint likelihood. We note that this can be generalized to an arbitrary number of lensed images and we refer the reader to \citep{Lo:2021nae} for more details. To obtain the posterior distribution over the parameters describing the joint likelihood function we use Bayes theorem and defer the details of our choice for the prior distribution to Section \ref{sec:results}.

\section{Results}\label{sec:results}
We perform joint parameter estimation to estimate the posterior distribution on the parameters defined through the strong lensing joint likelihood function as defined in Eq.~\eqref{eq:likelihood}. As an example, we simulate a pair of lensed GW images from a non-spinning binary black hole merger with the following intrinsic parameters: $m_1^{\text{det}} = 36 M_\odot$, $m_2^{\text{det}} = 29 M_\odot$ and $a_1 = a_2 = 0$. The extrinsic parameters for the simulated system are $\psi=2.659$, $\phi_c=2.9$, $\alpha=1.375$, $\delta=-1.2108$ and $\iota=\pi/4$. We have chosen a sky location for the merger consistent with the beam pattern functions as shown in the right panel of Fig. \ref{fig:antenna} and an inclination angle close to the value of $\iota_{\text{opt}}$ in order to not suppress the extra polarization modes through the inclination dependence introduce via the $g_A(\iota)$ factors.

As discussed in \ref{sec:pe}, we sample over the observed distance to the first event $D^{(1)}_{\text{obs}}$ and the relative amplification factor for the pair $\mu$. The first lensed pair has $D^{(1)}_{\text{obs}} = 1000 \ \text{Mpc}$ and $\mu=2$ (corresponding to $D^{(2)}_{\text{obs}} = 500 \ \text{Mpc}$) with a time delay $\Delta t = 6 \ \text{hours}$. We set the Morse index for both images to zero (both Type-I) for simplicity. For the polarization mode amplitudes we choose, $\epsilon_+ = \epsilon_\times = 0.35$, $\epsilon_x = \epsilon_y = 0.15$ and $\epsilon_s=0.05$. 

\begin{figure}[htb]
\includegraphics[width=0.47\textwidth]{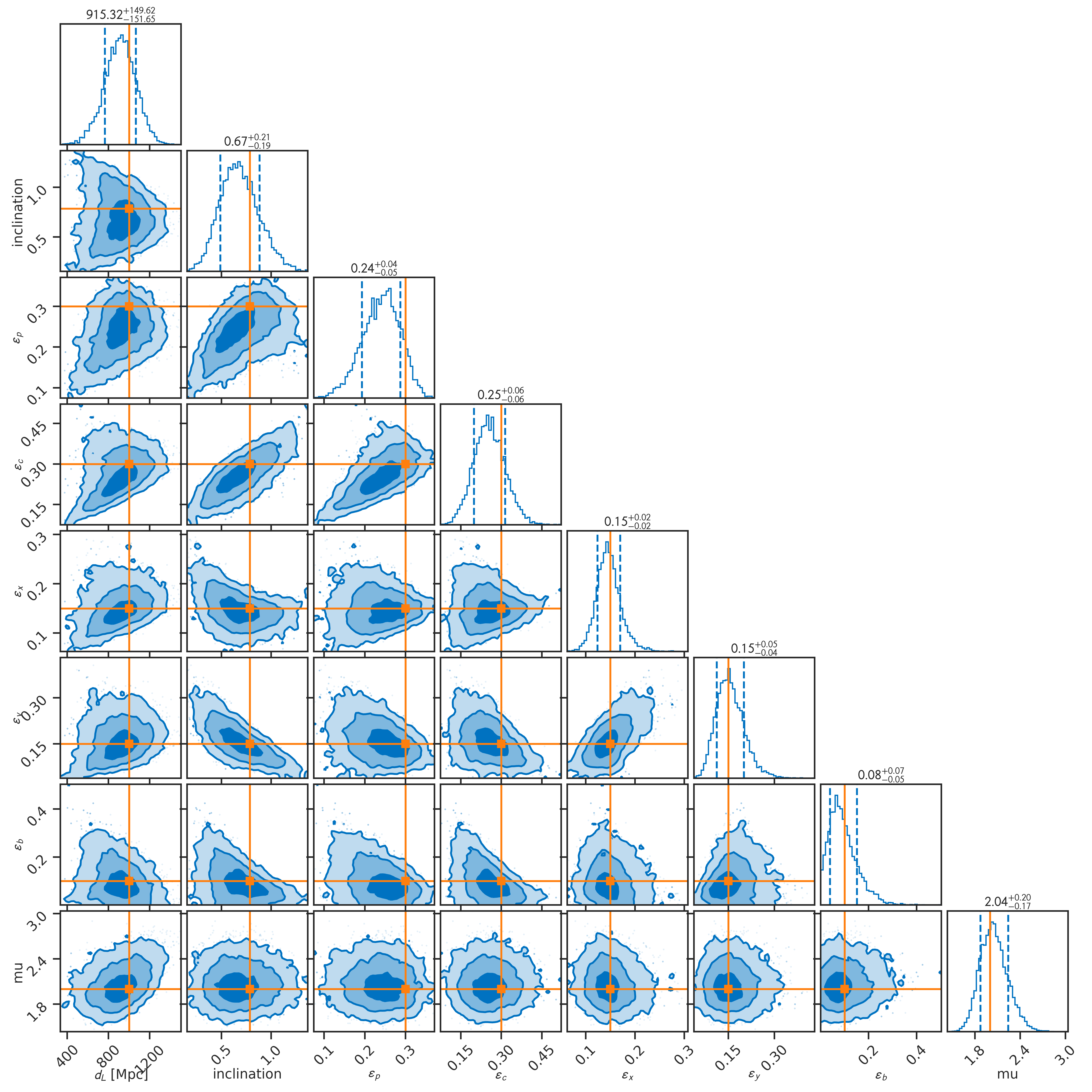}
\caption{Marginalized posterior distribution for the pair of lensed images as described in \ref{sec:results} observed by four detectors (HLVK) on the relative amplitudes for each polarization mode, inclination angle, relative magnification factor and the observed distance of the first image. The simulated system has $\epsilon_+ = \epsilon_\times = 0.35$, $\epsilon_x = \epsilon_y = 0.15$ and $\epsilon_s=0.05$ for the polarization mode amplitude, $D^{(1)}_{\text{obs}} = 1000 \ \text{Mpc}$, $\mu=2$ and $\iota=\pi/4$ (shown in orange) with a relative time delay of six hours. }
\label{Fig:postHLVK_amp}
\end{figure}

We consider two examples, a 2-detector network composed of LIGO Hanford and LIGO Livingston (HL) and a 4-detector network, with Advanced Virgo and KAGRA as additional detectors (HLVK), all at their corresponding design sensitivities. We generate the GW waveform using the \texttt{TaylorF2}\cite{Damour:2000zb} waveform model for simplicity and inject the two lensed GW signals into simulated data streams with Gaussian noise and sample over the strong lensing joint likelihood using \texttt{Bilby} \citep{Ashton:2018jfp,Romero-Shaw:2020owr}. For the 4-detector network, we show in Fig. \ref{Fig:postHLVK_amp} the marginalized posterior distribution on the relative polarization mode amplitudes, inclination angle, relative magnification factor and the observed distance to the first image (See Appendix \ref{sec:fullpe} for our prior choices as well as full parameter estimation results in Fig. \ref{Fig:postHLVK} for the HLVK case and in Fig. \ref{Fig:postHL} for the HL case). It is evident from the posterior distribution shown in Fig. \ref{Fig:postHLVK_amp} that the relative polarization mode amplitudes can be measured with a single pair of lensed events using a 4-detector network at design sensitivity. For the example with a 2-detector network observing the same system, the polarization mode amplitudes cannot be fully constrained due to the lack of linearly independent detectors (in principle four but both the Hanford and Livingston detectors are nearly co-aligned). 

\section{Discussion}
In this work we have performed Bayesian joint parameter estimation on pairs of strongly lensed GW events in order to constrain the relative amplitudes for alternative polarization modes using simulated data. We have used a simplified signal model as a proxy for the signal morphology for the additional polarization modes, and have also made sure to include the expected inclination angle dependence for each mode for GWs emitted by a merging binary. We have shown that the relative amplitudes as well as the amplitude relevant parameters such as the observed distance, inclination angle and relative amplification factor for the lensed pair can be measured, since the additional data from the same astrophysical system provides enough independent detectors to measure the aforementioned parameters. 

Strongly lensed pairs of GW signals for binary black hole mergers are expected to be detected as early as O4 but more likely in O5. Once a confident detection has been established, the joint parameter estimation framework described in this work can be applied to a real lensed pair of GW signals. However, we do mention that a proper treatment of real GW data will involve the strong lensing joint likelihood with a model independent framework to describe the GW signal morphology as explored in \cite{Chatziioannou:2021mij} which used \texttt{bayeswave} to model the GW signal morphology using sine gaussians. Given that, the results of this paper can be seen as being slightly pessimistic than what they would be if any alternative polarization modes are present in the data with significantly different signal morphology. The varying morphology should allow for the relative mode amplitude degeneracy to be broken, however, using a specific modified gravity model that predicts additional polarization modes for the Bayesian inference would make the results model dependent.

\section*{Acknowledgements} 
The author would like to thank Virginia d'Emilio, Jolien Creighton, Soichiro Morisaki and Anarya Ray for useful comments and feedback throughout this work. IMH is supported by the NSF Graduate Research Fellowship Program under grant DGE-17247915. This work was supported by NSF awards PHY-1912649. The author is grateful for computational resources provided by the Leonard E Parker Center for Gravitation, Cosmology and Astrophysics at the University of Wisconsin-Milwaukee. We thank LIGO and Virgo Collaboration for providing the data for this work. This research has made use of data, software and/or web tools obtained from the Gravitational Wave Open Science Center (https://www.gw-openscience.org/), a service of LIGO Laboratory, the LIGO Scientific Collaboration and the Virgo Collaboration. LIGO Laboratory and Advanced LIGO are funded by the United States National Science Foundation (NSF) as well as the Science and Technology Facilities Council (STFC) of the United Kingdom, the Max-Planck-Society (MPS), and the State of Niedersachsen/Germany for support of the construction of Advanced LIGO and construction and operation of the GEO600 detector. Additional support for Advanced LIGO was provided by the Australian Research Council. Virgo is funded, through the European Gravitational Observatory (EGO), by the French Centre National de Recherche Scientifique (CNRS), the Italian Istituto Nazionale di Fisica Nucleare (INFN) and the Dutch Nikhef, with contributions by institutions from Belgium, Germany, Greece, Hungary, Ireland, Japan, Monaco, Poland, Portugal, Spain. This material is based upon work supported by NSF's LIGO Laboratory which is a major facility fully funded by the National Science Foundation. This article has been assigned LIGO document number LIGO-P2200329.

\appendix
\section{Full Posterior Distributions}
\label{sec:fullpe}
We provide the full posterior distributions for the cases investigated in Sec. \ref{sec:results} for completeness. 

\onecolumngrid
\begin{figure}[htb]
\centering
\includegraphics[width=0.69\textwidth]{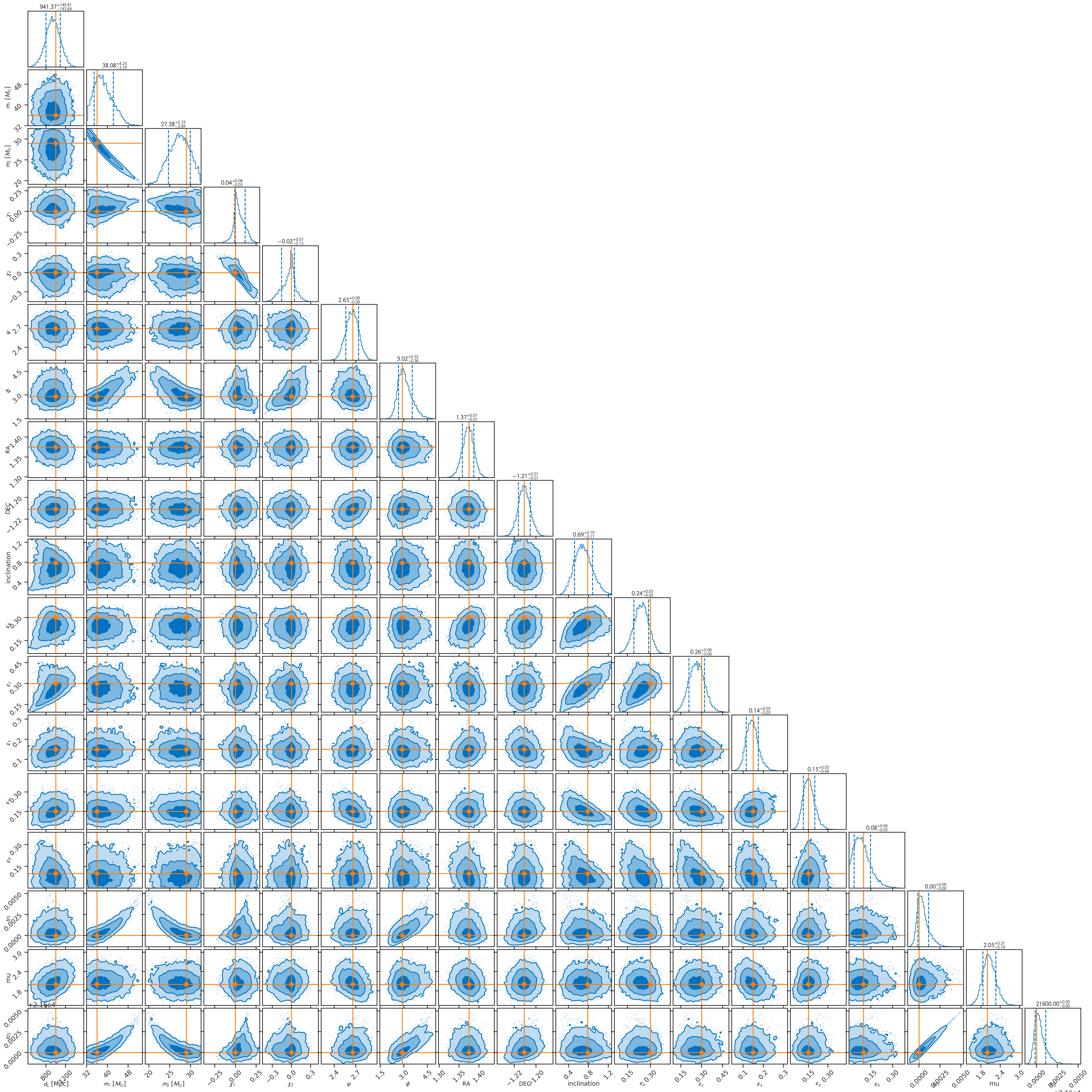}
\caption{Posterior distribution on the pair of lensed images as described in Section \ref{sec:results} observed by four detectors (HLVK) with a relative time delay of six hours with polarization mode amplitude of $\epsilon_+ = \epsilon_\times = 0.35$, $\epsilon_x = \epsilon_y = 0.15$ and $\epsilon_s=0.05$. }
\label{Fig:postHLVK}
\end{figure}

\newpage
\onecolumngrid
\begin{figure}[htb]
\centering
\includegraphics[width=0.69\textwidth]{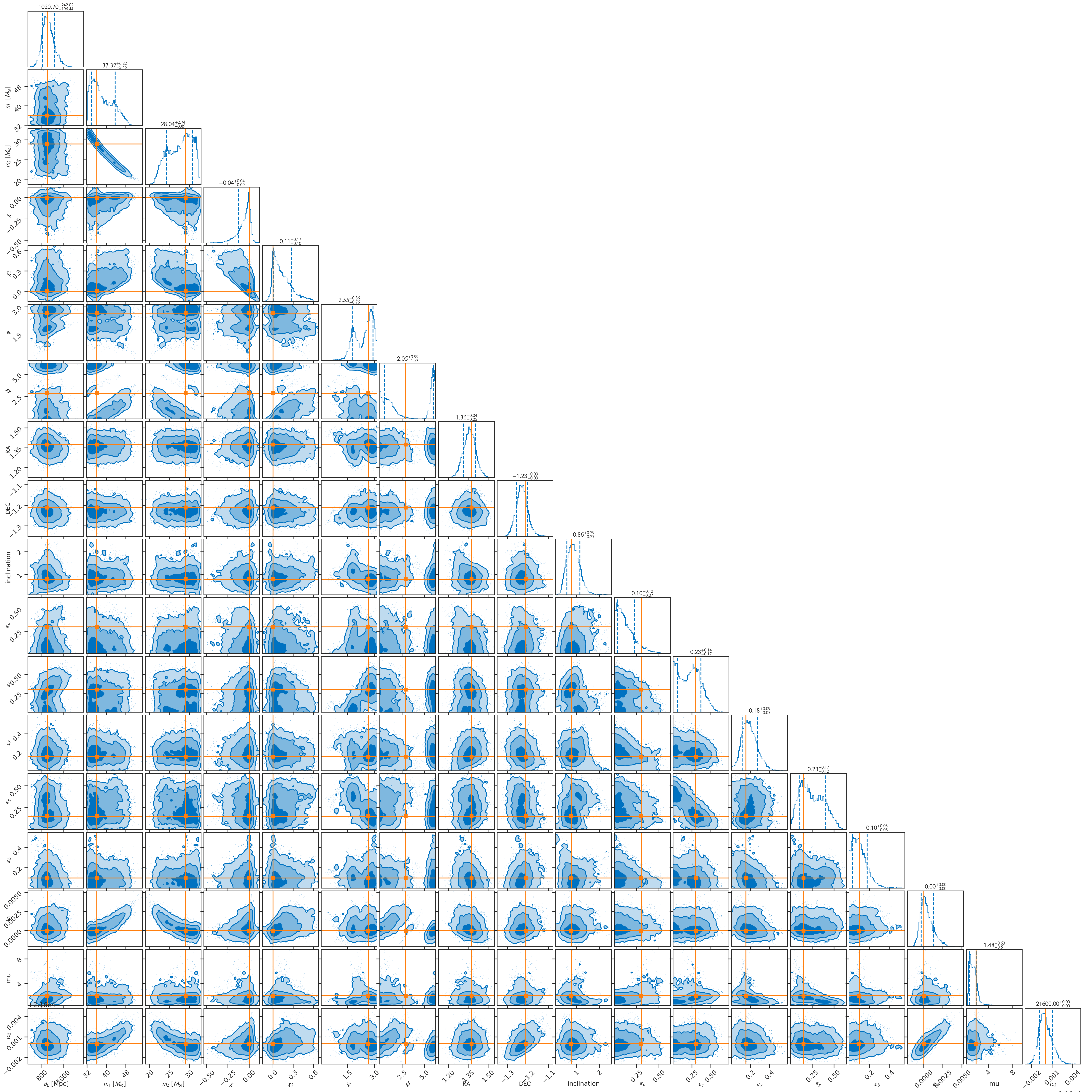}
\caption{Posterior distribution on the pair of lensed images as described in Section \ref{sec:results} observed by two detectors (HL). }
\label{Fig:postHL}
\end{figure}

\newpage
\bibliography{references}

\end{document}